# THE GENERATION OF VARIABLE POLARIZATION STATES IN TERAWATT X-RAY FREE-ELECTRON LASERS

H.P. Freund[1,2] and P.G. O'Shea[1,3]

[1]Institute for Research in Electronics and Applied Physics, University of Maryland, College Park, MD 20742, USA
[2]NoVa Physical Science and Simulations, Vienna, Virginia 22182, USA
[3]Department of Electrical Engineering, University of Maryland, College Park, Maryland 20742, USA

Terawatt x-ray free-electron lasers (XFELs) represent the frontier in further development of x-ray sources and require high current densities with strong transverse focusing. In this paper, we investigate the implications/potentialities of TW XFELs on the generation of harmonics at still shorter wavelengths and higher photon energies with variable polarization. The simulations indicate that significant power levels are possible at high harmonics of the XFEL resonance and that these XFELs can be an important coherent source of hard x-rays through the gamma ray spectrum. For this purpose, we use the MINERVA simulation code which self-consistently includes harmonic generation. Both helical and planar undulators are discussed in which the fundamental is at 1.5 Å and study the associated harmonic generation. While tapered undulators are needed to reach TW powers at the fundamental, the taper does not enhance the harmonics because the taper must start before saturation of the fundamental, while the harmonics saturate before this point is reached. Nevertheless, the harmonics reach substantial powers. Simulations indicate that, for the parameters under consideration, peak powers of the order of 180 MW are possible at the fifth harmonic with a photon energy of about 41 keV and still high harmonics may also be generated at substantial powers. Such high harmonic powers are certain to enable a host of enhanced applications.

## I. INTRODUCTION

The ability to create and manipulate x-ray photon polarization states is of great importance for many applications in materials characterization and imaging, e.g for magnetic [1] and quantum materials [2], and in molecular and biophysics [3,4]. In this paper, we show that x-ray Free-Electron Laser facilities (XFELs) that are currently operating or are under construction worldwide [5-13] can be capable of producing powerful photon beams with variable elliptical polarization.

The first operational XFEL was the Linac Coherent Light Source (LCLS) at the Stanford Linear Accelerator Center [5] which produced 20 GW/83 fs pulses at a 1.5 Å wavelength with a repetition rate of 120 Hz operating in self-amplified spontaneous emission (SASE) mode. As such, there is active interest in the study of novel configurations to improve the output power, spectral stability, spectral purity, temporal characteristics, self-seeding, and etc. Of particular interest to us in this paper, simulations have been conducted recently to study the conditions necessary for the generation of terawatt output powers [14-16]. Achieving TW power levels will most likely require step-tapered, superconducting undulators (SCUs).

Substantial progress has been made in the development of SCUs [17-24]. Superconducting undulators are under development for all current undulator configurations including planar, helical and APPLE-II undulator designs. In this paper, we are particularly interested in superconducting APPLE-II (SCAPE) undulators [24] under development at Argonne National Laboratory since this design facilitates the development of TW XFELs with the capability of generating arbitrary polarization states simply by retuning the APPLE-II undulators.

In the present work, we consider variable polarization TW generation in APPLE-II undulator lines using the MINERVA simulation code [15,25-28] which self-consistently treats the *x*- and *y*-components of the optical field independently and follows the evolution of the polarization in any arbitrary undulator configuration. It was found previously [15] that TW generation required extreme focusing of the beam with current densities in excess of 1 GA/cm$^2$. Since undulators provide only weak focusing, this implies that an extremely strong focusing/defocusing (FODO) lattice composed of quadrupoles with short-period separation (short FODO cells) is required. Since the undulators are located in the intervals between the quadrupoles, this implies that a relatively large number of short undulators ($\approx$ 1 m) is required. The results of these simulations indicate that TW power levels can be generated with polarizations ranging from linear through circular polarizations of the undulators.

The organization of the paper is as follows. The geometrical configuration and the numerical formulation are described in Sec. 2. Simulations of the configuration at the fundamental resonance are presented in Sec. 3. This includes the variations in performance as the polarization of the APPLE-II undulators vary. Harmonic generation in a TW XFEL has been discussed previously [29] where it was demonstrated that higher harmonic powers are generated in a planar undulators than in helical undulators. The variation in the generation of harmonics with the polarization of the APPLE-II undulators is discussed in Sec. 4. A summary and discussion is presented in Sec. 5.

Corresponding author: freundh0523@gmail.com

## II. THE GEOMETRICAL CONFIGURATION AND THE NUMERICAL FORMULATION

The configuration under consideration consists of an alternating sequence of undulators and quadrupoles. Since extremely strong focusing is necessary, the FODO cell length is 2.16 m. The overall geometry is shown in Fig. 1 which includes a segment consisting of two undulators and quadrupoles. The short distance between quadrupoles imposes the necessity of relatively short undulators, and we use 46 period undulators with a period of 2.0 cm for an overall length of 92 cm. each undulator includes one period each ($N_{trans} = 1$) for the entry and exit transitions; hence, the are 44 periods of uniform field. The quadrupoles are placed in the center of the gaps between the undulators.

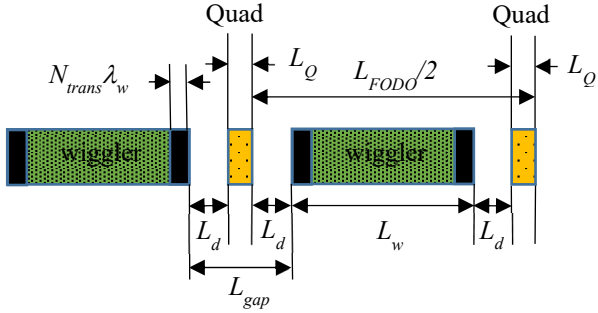

Fig. 1: Illustration of one FODO cell of the undulator line.

| Undulators: APPLE-II | |
| --- | --- |
| Period, $\lambda_w$ | 2.0 cm |
| Length, $L_w$ | 96 cm |
| Entry/Exit transitions | 1 period |
| **Quadrupoles** | |
| Field Gradient | 26.4 kG/cm |
| Length, $L_Q$ | 7.4 cm |
| FODO Cell Length, $L_{FODO}$ | 2.16 m |
| **Separation Distances** | |
| Drift Length, $L_d$ | 4.3 cm |

Table 1: Undulator, FODO cell parameters.

The electron beam model used a kinetic energy of 13.2 GeV, a current of 4000 A with a flat-top temporal profile with a duration of 24 fs for a bunch charge of 96 pC, an emittance of 0.3 μm and an rms energy spread of 0.01%. Since the FODO lattice dominates the beam propagation, the Twiss parameters used to match the beam into the undulator line were held fixed for all the tunings of the APPLE-II undulators and corresponded to an initial rms x-dimension of 8.03 μm with a Twiss $\alpha_x = 1.29$ and an initial rms y-dimension of 6.22 μm with a Twiss $\alpha_y = -0.78$. These parameters are summarized in Table 2.

Beam propagation is relatively insensitive to the undulator parameters in the strong focusing FODO lattice, and a plot of the evolution of the beam envelopes in the x- and y-directions for a representative undulator is shown in Fig. 2 over a distance of 100 m.

| Kinetic Energy | 13.2 GeV |
| --- | --- |
| Current | 4000 A |
| Pulse Duration (flat-top) | 24 fs |
| Bunch Charge | 96 pC |
| Normalized Emittance | 0.3 μm |
| rms Energy Spread | 0.01% |
| rms x-Dimension | 8.03 μm |
| Twiss $\alpha_x$ | 1.29 |
| rms y-Dimension | 6.22 μm |
| Twiss $\alpha_y$ | −0.78 |

Table 2: Electron beam parameters.

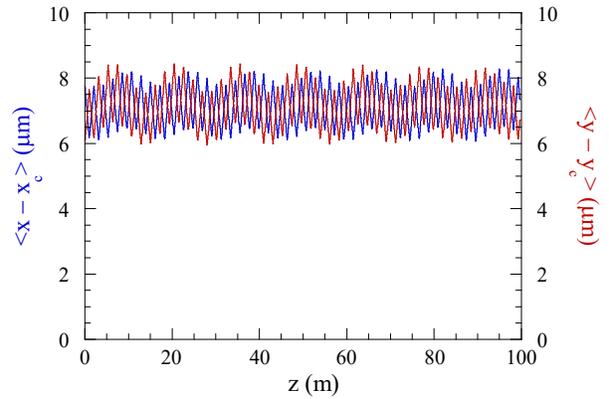

Fig. 2: Evolution of the beam envelope.

We choose to operate at a fixed wavelength of 1.5 Å. In an APPLE-II undulator, the peak on-axis magnetic field will change as the polarization is changed and this will retune the FEL resonance. However, for computational simplicity, we will vary the undulator field as we retune the polarization in such a way as to hold the resonant wavelength fixed.

MINERVA integrate the orbit equations using an approximate analytic field model for an APPLE-II undulator [30] which is a superposition of two planar undulators where one is shifted in phase ($\phi$) as follows

$$\mathbf{B}_w = B_w \left( \sin(k_w z + \phi) - \frac{\cos(k_w z + \phi)}{k_w B_w} \frac{dB_w}{dz} \right) \hat{e}_x \cosh k_w x$$
$$+ B_w \left( \sin k_w z - \frac{\cos k_w z}{k_w B_w} \frac{dB_w}{dz} \right) \hat{e}_y \cosh k_w y$$
$$+ B_w \hat{e}_z \sinh k_w y \left[ \cos(k_w z + \phi) + \cos k_w z \right], \quad (1)$$

where $k_w$ is the undulator wavenumber and the ellipticity, $u_e$, varies over the range $0 \leq u_e \leq 1$ from planar polarization ($u_e = 0$) to circular polarization ($u_e = 1$) and is given by

$$u_e = \begin{cases} \dfrac{1-\cos\phi}{1+\cos\phi} & ; 0 \leq \phi \leq \pi/2 \\ \dfrac{1+\cos\phi}{1-\cos\phi} & ; \pi/2 < \phi \leq \pi \end{cases}. \quad (2)$$

The resonance condition varies with the ellipticity via

$$\lambda_{res} = \frac{\lambda_w}{2\gamma^2}\left[1 + \left(1 + u_e^2\right)\frac{K^2}{2}\right], \quad (3)$$

where $K$ is the undulator strength parameter and which we vary as the ellipticity is tuned so that the resonant wavelength is held constant.

The variation in the peak on-axis magnetic field versus the ellipticity of the APPLE-II undulators to achieve a resonance at 1.5 Å is shown in Fig. 3.

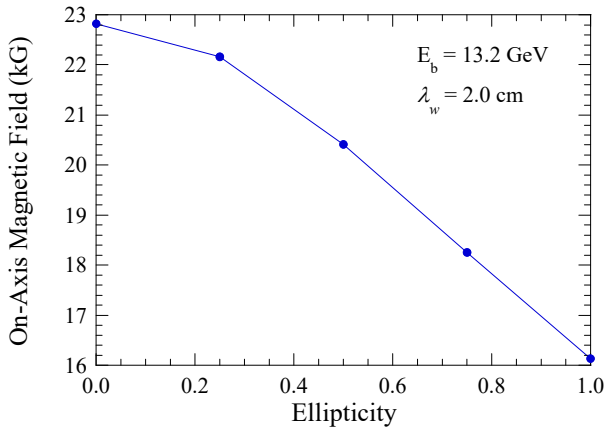

Fig. 3: Variation in the on-axis magnetic field with the ellipticity.

MINERVA is a three-dimensional, time-dependent nonlinear formulation for modeling amplifier, oscillator, and self-amplified spontaneous emission (SASE) configurations. MINERVA uses the Slowly-Varying Envelope Approximation (SVEA) and the optical field is represented by a superposition of Gauss-Hermite modes. The field equations are averaged over the rapid sinusoidal time scale and, thereby, reduced to equations describing the evolution of the slowly-varying amplitude and phase. The $x$- and $y$-components of the field are integrated independently; hence, MINERVA is capable of self-consistently simulating an undulator line composed of a variety of different polarizations. Time-dependence is treated using a breakdown of the electron bunch and the optical pulse into temporal *slices* each of which is one wave period in duration. The optical *slices* are allowed to slip ahead of the electron slices. MINERVA integrates each electron and optical *slice* from $z \rightarrow z + \Delta z$ and the appropriate amount of slippage can be applied after each step or after an arbitrary number of steps by interpolation.

Particle orbits are integrated using the full Lorentz force equations in the complete optical and magnetostatic fields (undulators, quadrupoles and dipoles). It is important to remark that the use of the full Lorentz orbit analysis allows MINERVA to self-consistently treat both the entry/exit tapers of undulators, and harmonic generation. Thus, MINERVA self-consistently tracks the particle distribution and optical field through the undulator line and includes optical guiding and diffraction as well as the associated phase advance of the optical field relative to the electrons throughout.

The dynamical equations for the particles and field amplitudes are integrated simultaneously in MINERVA using a 4$^{th}$ order Runge-Kutta algorithm; hence, it is possible to change the step size on the fly. In the simulation of the undulator line/FODO lattice, MINERVA will take an integer number of steps through an undulator (typically 20 or 30 steps per undulator period). Upon leaving the undulator, MINERVA will then determine the distance to the next magnetic element such as a quadrupole, and change the step size so as to reach the quadrupole in an integer number of steps (typically a longer step size through a field-free-region). MINERVA will then change (shrink) the step size again to take an integer number of steps through the quadrupole. Upon leaving the quadrupole, MINERVA will determine the distance to the next magnetic element (an undulator) and lengthen the step size through the field-free-region until it reaches the undulator in an integer number of steps. At that point it will again shrink the step size to that chosen for the undulator. Throughout this integration procedure, MINERVA self-consistently tracks the relative phase advance of the optical field relative to the electrons. In this way, MINERVA accurately simulates the entire undulator line/FODO lattice.

### III. TW GENERATION AT THE FUNDAMENTAL

It will be necessary to employ a tapered undulator line in order to achieve terawatt power levels. To this end, we consider a 100-m long undulator line/FODO lattice, with a step-taper from undulator to undulator. The efficacy of tapered undulators depends upon both the slop of the taper and the start-taper point. The optimization of the start-taper point in a segmented undulator line is relatively coarse since the optimal start-taper point may be within any given undulator and it is not possible to optimize more effectively than choosing a specific undulator as the start-taper point. It is important to remark that the undulator line consists of 91 undulators and the optimal taper begins at about the 30$^{th}$ – 35$^{th}$ undulator. This means that there are approximately 60 step-tapered undulators before the end of the undulator line and a small variation in the optimal start-taper point can give rise to a relatively large variation in the output energy. Variations in the optimal start-taper point are also expected since we are considering start-up due to SASE. As a result, fluctuations are found in the output power versus the ellipticity of the undulators due to this limitation in the optimization procedure.

In order to determine the output pulse energy in a SASE configuration, we determine the average pulse energy over an ensemble of simulation runs with different noise seeds.

Typically, this requires about 15 different noise seeds to achieve convergence. For the cases under consideration the level of fluctuations due to SASE varies with the ellipticity but is in the range of 3 – 6 percent. An example of the evolution of the average pulse energy versus distance through the undulator line is shown in Fig. 4 for an undulator ellipticity of 0.50 in which the average pulse energy at the end of the undulator line is about 4.6 mJ.

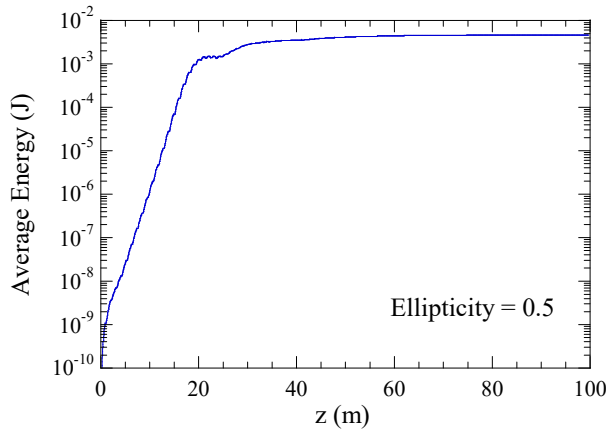

Fig. 4: the evolution of the average output pulse energy versus the ellipticity of the undulator line.

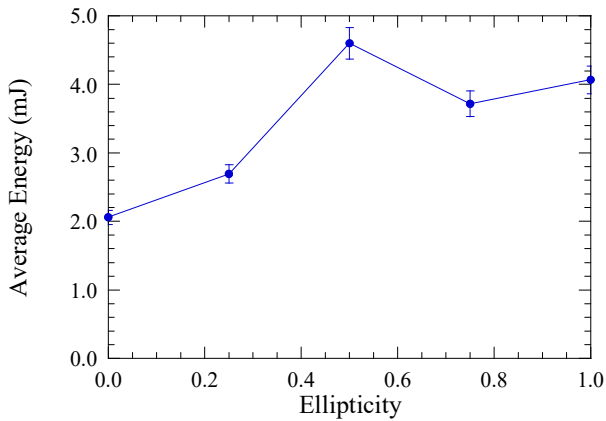

Fig. 5: Variation in the average output power versus the ellipticity of the undulators. The error bars indicate the level of SASE fluctuations from shot-to-shot.

The output pulse energy tends to increase with increases in the ellipticity of the undulators. This is expected since the interaction is stronger in a helical undulator than in a planar undulator. This is shown in Fig. 5 in which we plot the average output energy versus the ellipticity of the undulators. We attribute the local maximum at the ellipticity of 0.50 to variations in the start-taper point due to the SASE process and the coarseness of the step-taper and this particular result may vary with detailed changes in the configuration under consideration.

The variation in the eccentricity versus the ellipticity of the undulators, using the Stokes diagnostic [28], is shown in Fig. 6. As expected, the polarization at the undulator exit exhibits purely linear polarization when the Apple-II undulators are tuned for planar polarization with an ellipticity = 0.0. The degree of elliptical polarization increases, and the eccentricity decreases, as the ellipticity of the undulators increases in a nearly linear fashion down to an eccentricity of approximately 0.14 when the ellipticity = 1.0. This indicates that the Apple-II undulators do not produce purely circular polarization.

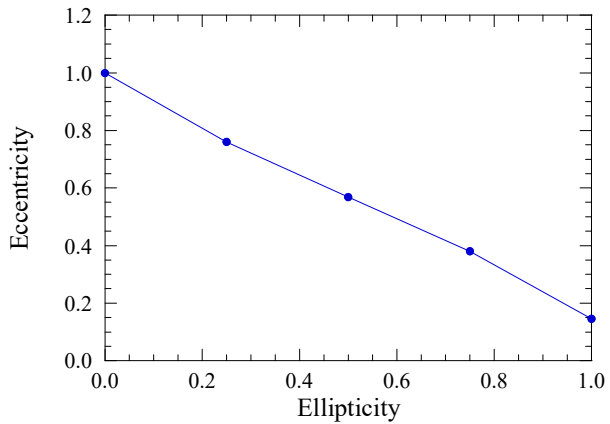

Fig.6: Plot of the eccentricity with variation in the undulator ellipticity.

The beam energy and undulators have been tuned to wavelengths close to 1.5 Å. The output spectra at ellipticities of 0.0, 0.5, and 1.0 are shown in Figs. 7-9 respectively. Each of these spectra exhibit the typical spikiness associated with the SASE mechanism. This spikiness can be eliminated by employing a self-seeding scheme; however, this is a trade-off because the self-seeding would introduce enhanced shot-to-shot fluctuations. Nevertheless, the relative linewidth is small. The rms linewidth for an ellipticity of 0.0, as shown in Fig. 7, is 0.12% and is higher than that found for the cases of elliptical polarization for which the rms relative linewidths were of the order of 0.066%. This is shown in Fig. 10.

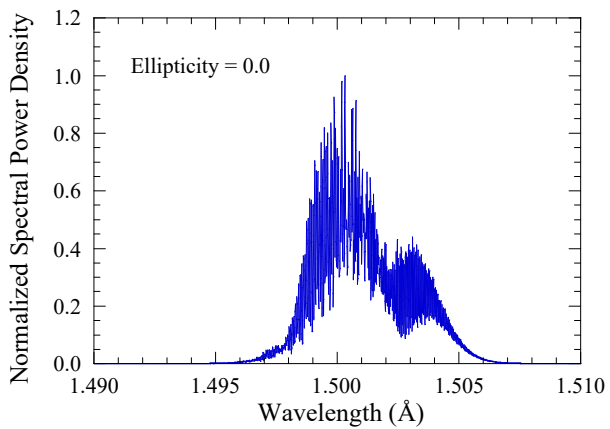

Fig. 7: The output spectrum corresponding to an ellipticity of 0.0.

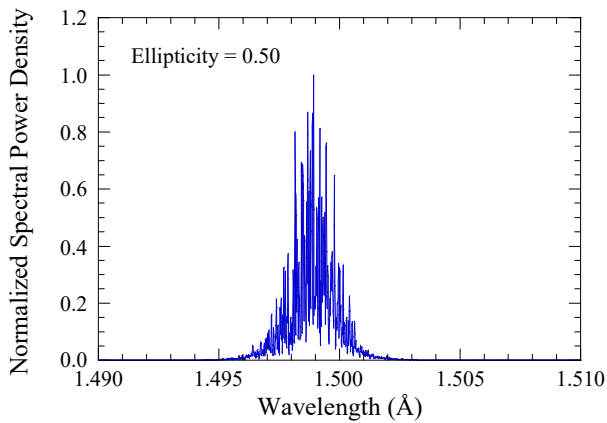

Fig. 8: The output spectrum corresponding to an ellipticity of 0.05.

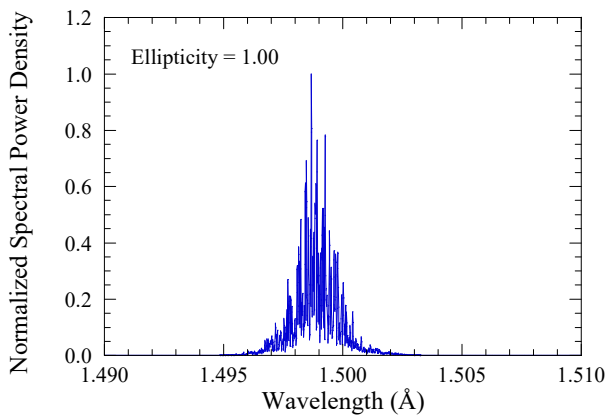

Fig. 9: The output spectrum corresponding to an ellipticity of 1.0.

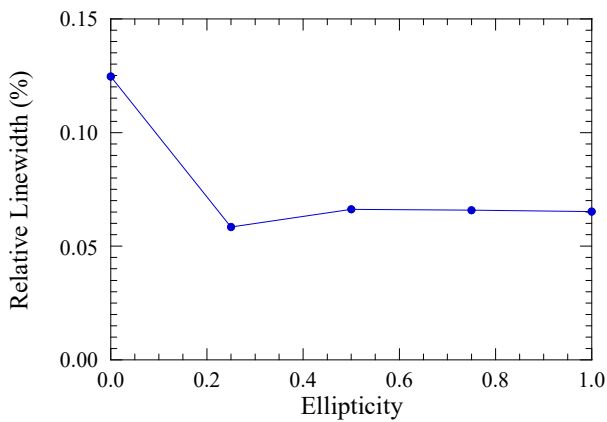

Fig. 10: Variation in the eccentricity versus the ellipticity.

**IV. HARMONIC GENERATION**

It was explicitly demonstrated that harmonic generation in a TW XFEL configuration is stronger for planar undulators than for helical undulators [29]. In this paper, we address the question of how the relative strength of harmonic generation varies as the APPLE-II undulators are tuned from pure planar polarization to pure helical polarization.

It is well known that harmonic generation in planar undulators is driven by oscillations in the axial velocity which favor the generation of odd harmonics while harmonic generation in helical undulators is due to a phase resonance whereby the $h^{th}$ harmonic represents a phase resonance with circularly polarized waves that rotate through $2\pi h$ radians per undulator period irrespective of the harmonic number [30]. As such, the strength of the harmonic generation in planar undulators is sensitive to the wiggler strength and increases as the undulator strength parameter, $K$, increases. However, because it is a phase resonance in helical undulators, the harmonic intensities are relatively insensitive to the undulator strength parameter. It is expected, therefore, that harmonic generation will be through a superposition of these mechanisms as the ellipticity is varied.

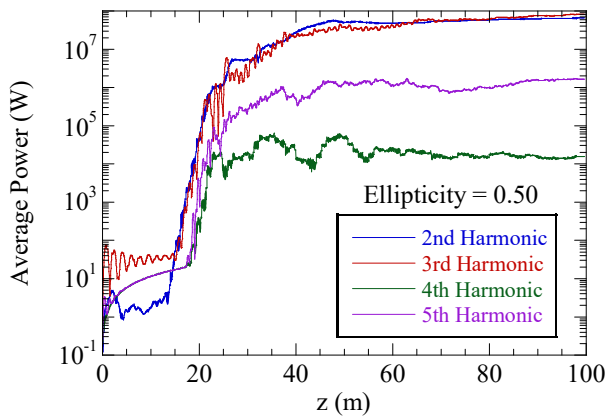

Fig. 11: Evolution of the powers in the $2^{nd} - 5^{th}$ harmonics for the case of the undulator ellipticity of 0.50.

This latter point is shown in Fig. 11 where we plot the evolution of the $2^{nd} - 5^{th}$ harmonics for undulator ellipticities of 0.50. The harmonic powers start to grow slowly until the fundamental reaches significant power levels of about 10 GW after about 20 m (fundamental not shown) at which point rapid exponential growth is found where the growth rate varies inversely with the harmonic number [31]. It is generally expected that the $3^{rd}$ harmonic will be excited more strongly than the $2^{nd}$ harmonic in purely planar undulators while the opposite will be the case in purely helical undulators. In the case shown in Fig. 11, the ellipticity is midway between these two extremes and the $2^{nd}$ and $3^{rd}$ harmonics are excited with nearly equal intensities. As pointed out previously [29], the harmonics are relatively unaffected by the taper since they saturate at a point prior to that of the fundamental.

Fluctuations in the harmonics due to the SASE process have been determined by simulations of ensembles of 15 – 25 different noise seeds and are found to be greater than those in the fundamental and reach anywhere from 10 – 25 percent depending upon the ellipticity and harmonic number

It should be mentioned that in order to ensure that the phase space is accurately resolved for the 2nd the 5th harmonics, we have included almost 500,00 particles in the simulation.

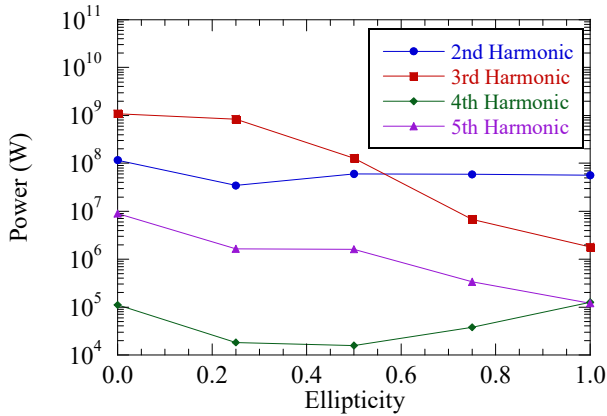

Fig. 12: Variation in the output power versus the ellipticity of the undulators.

A summary plot of the strength of harmonic generation as the ellipticity of the undulators is varied from 0.0 – 1.0 is shown in Fig. 12 where it is seen, as expected, that the 3rd harmonic is excited more strongly than the 2nd harmonic for planar undulators while the opposite is found for helical undulators.

## V. SUMMARY AND DISCUSSION

Important applications of high power XFELs exist for a variety of different polarization states. With the advent of superconducting APPLE-II undulators [24] which can tune the polarization from linear through helical it is possible to design an XFEL which can achieve terawatt-class operation across the entire range of ellipticities. In this paper, we have considered one such configuration of APPLE-II undulators which are step-tapered and immersed in a strongly focusing FODO lattice that can support current densities of the order of 2 GA/cm$^2$. Using this configuration, we have studied how the performance varies with the ellipticity of the undulators for both the fundamental ad the 2nd – 5th harmonics.

## ACKNOWLEDGMENTS

This work was supported by the United States Department of Energy under contract DE-SC0024397.